\documentclass[a4paper,superscriptaddress,twocolumn,english]{revtex4}
\usepackage{bm}
\usepackage{amsmath}
\usepackage{amssymb}
\usepackage{graphicx}
\usepackage{epsfig}
\usepackage{epstopdf}
\usepackage{hyperref}

\usepackage{dsfont}
\usepackage{dcolumn}

\DeclareMathOperator{\Tr}{Tr}
\DeclareMathOperator{\Const}{const}
\newcommand{\hmm}[1]{#1\nobreak\discretionary{}{\hbox{\ensuremath{#1}}}{}}

\begin{document}

\title{Data-pattern tomography of entangled states}

\author{Vadim Reut}
\affiliation{
\small{B. I. Stepanov Institute of Physics, National Academy of Science of Belarus, Nezavisimosti Ave. 68, Minsk 220072 Belarus}\\}
\affiliation{\small{Department of Theoretical Physics and Astrophysics, Belarusian State University, Nezavisimosty Ave. 4, Minsk 220030 Belarus}\\}
\author{Alexander Mikhalychev}
\affiliation{
\small{B. I. Stepanov Institute of Physics, National Academy of Science of Belarus, Nezavisimosti Ave. 68, Minsk 220072 Belarus}\\}
\author{Dmitri Mogilevtsev}
\affiliation{
\small{B. I. Stepanov Institute of Physics, National Academy of Science of Belarus, Nezavisimosti Ave. 68, Minsk 220072 Belarus}\\}

\date{\today}

\begin{abstract}
We discuss the data-pattern tomography for reconstruction of entangled states of light. We show that for a moderate number of probe coherent states it is possible to achieve high accuracy of representation not only for single-mode states but also for two-mode entangled states.  We analyze the stability of these representations to the noise and demonstrate the conservation of the purity and entanglement. Simulating the probe and signal measurements, we show that systematic error inherent for representation of realistic signal response with finite sets of responses from probe states still allows one to infer reliably the signal states preserving entanglement.
\end{abstract}

\maketitle


\section{\label{sec:intro}Introduction}

Quantum tomography as a way of inferring a quantum state is potentially the most precise measuring tool available to a physicist \cite{paris0,nielsen_chuang,yongsiahbook}. However, this tool requires rather precise tuning. One needs to know characteristics of the measurement setup which necessarily involves a calibration of it. Generally, it is quite a nontrivial task equivalent to quantum process tomography of the detecting system. Provided that one can describe the setup with the few-parameter model (such as efficiency and dark count rate of detectors, etc.), it is possible in some cases to perform the calibration without a complete set of known probes by trading some information about the probe for knowledge about the detector. For example, with a twin-photon state, one can find the absolute value of the detecting setup efficiency ~\cite{klyshko-1980,malygin-1981}; entanglement also makes possible ``self-testing'' or ``blind tomography''~\cite{scarani-2012,scarani-2014}. Trading knowledge of probes (preferably of the most general nature, such as Gaussianity) for information about the measurement gave rise to the concept of self-sufficient, or self-calibrating tomography \cite{mogilevtsev-2009,mogilevtsev-2010calibration,branczyk-2012,mogilevtsev-2012,stark2016}.

However, there is a possibility of skipping the calibration stage altogether. This possibility is given by  the data-pattern  tomography~\cite{mogilevtsev-2010qt,mogilevtsev-2013}. The idea of this method is somewhat similar to that of optical image analysis with a known optical response function~\cite{gonzalez-2002}. An observer measures responses (the data patterns) for a set of known quantum probe states and matches them with the response obtained from the unknown signal of interest. The data-pattern tomography can also be understood as a search for the optimal state estimator over the subspace that is spanned by the probe states. This approach is naturally insensitive to imperfections of the measurement setup since all device imperfections are automatically incorporated  into and accounted for by the measured data patterns. The data-pattern scheme was recently successfully realized with few-photon signals and coherent probes and was shown to be quite robust~\cite{cooper-2014,harder-2014}.

The efficiency of data-pattern tomography depends essentially on the choice of the basis set of probe states. It is highly desirable to use the smallest possible number of basis states.  If the observer believes that the signal state is very likely residing in some operator subspace, he or she can make use of this insight to define the set of probe states that spans this subspace for data-pattern reconstruction~\cite{mogilevtsev-2010qt,mogilevtsev-2013}. Naturally, the accuracy of the method depends crucially on the accuracy of the signal representation. Systematic error intrinsic in the method is unavoidably amplified in the process of the signal inference since both probe and signal patterns are subject to statistical errors.

Here we show that such an amplification of errors would not lead to breaking of essentially quantum features of the state. Such a fragile feature as the entanglement survives the inference procedure, and for sufficiently high accuracy of the representation the fidelity of the signal reconstructions for the large number of signal state copies tends to the values close to the fidelity given by the representation. Our simulation shows that this number of copies stands well within the region of experimental feasibility. Moreover, entanglement is not broken even for a comparatively low number of signal copies, when infidelity is several times higher than the infidelity of the representation (which is rather remarkable if one takes into account the main feature of the data-pattern approach: the density matrix of the signal is approximated by the mixture of nonorthogonal projectors). So the data-pattern scheme is quite feasible and reliable also for the reconstruction of multimodal states. Also, we show that the representation of the signal in terms of the classical probe basis is quite robust with respect to the noise of the representation weights.

The outline of the article is as follows. In Sec.~\ref{sec:basis} we review the basics of the data-pattern scheme and discuss the selection criteria of optimal basis sets. After that, we analyze different basis sets of coherent states for the representation of single-mode states as the initial simple problem and entangled double-mode states with a small average number of photons. Next, in Sec.~\ref{sec:analysis}, the quality of such expansions based on the sets with optimal parameters is discussed. We analyze the stability of this procedure to the noise and evaluate entanglement and the purity of the represented states. Last, but not least, in Sec.~\ref{sec:reconstruction} we present simulations of the procedure of data-pattern reconstruction using optimal sets of coherent projectors for single-mode and entangled double-mode states and demonstrate the survival of entanglement.

\section{\label{sec:basis}Optimal basis sets}
In this section, we consider the representation of optical quantum states based on the discrete basis set in data-pattern tomography. From a practical point of view, it is essential to find optimal basis sets with the minimum possible number of coherent projectors for the representation.  However, to our knowledge, the optimization of the basis choice has not yet been fully discussed and analyzed (here one can point to only two preliminary recent works \cite{mogilevtsev-2013,mogilevtsev-2014}). In experiments~\cite{cooper-2014,harder-2014} the reconstruction was done by considering only a set of probe states that was \textit{a priori} deemed sufficiently large (from $48$ to $150$). We analyze the applicability of different sets of probe states for data-pattern tomography. For the chosen basis set, the efficiency of the reconstruction could be enhanced using, for example, the adaptive Bayesian procedure \cite{holsby, straupe1,straupe2,mikhalychev-2015}.

\subsection{\label{sec:basis:principles}General principles}
\begin{figure}[!htb]
\includegraphics[width=8.6cm]{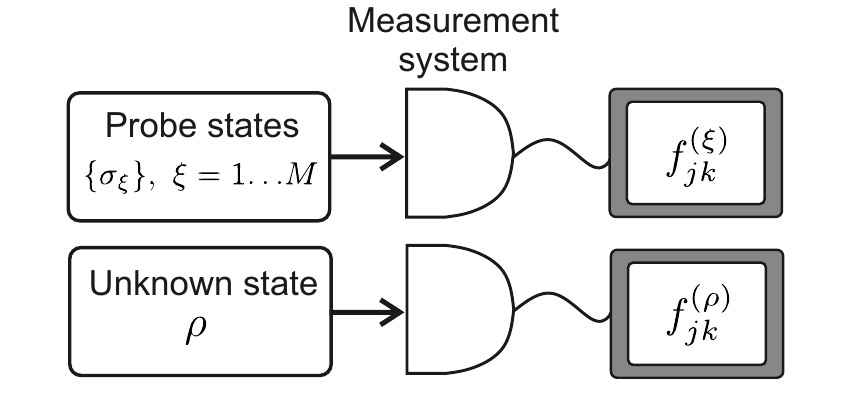}
\caption{Measurement scheme of data-pattern approach.}\label{fig:scheme}
\end{figure}

First of all, let us consider the general principles of the data-pattern scheme. We assume that there is an appropriately chosen finite set of probe states which can be described by the density operators $\sigma_\xi$, where $\xi=1,\dots{},M$. We would like to reconstruct the true signal state described by the density operator $\rho$. The key point of the discussion is the possibility to fit the signal $\rho$ with a mixture of probes,
\begin{equation}
\label{eq:expansion}
\rho\approx\rho^{Appr}=\sum_{\xi=1}^{M}x_{\xi}\sigma_{\xi},
\end{equation}
where $x_{\xi}$ are real coefficients. In order to proceed with the reconstruction, an observer carries out a number of some measurements on the unknown signal state $\rho$ and a predefined set of probe states $\sigma_{\xi}$. The outcome $k$ in the $j$-th measurement can be described by positive operator-valued measures (POVM) $\Pi_{jk}$,
\begin{eqnarray}
\label{eq:freqprob}
\begin{array}{lr}
p^{(\xi)}_{jk}=Tr(\Pi_{jk}\sigma_\xi),\\[3pt]
p^{(\rho)}_{jk}=Tr(\Pi_{jk}\rho),
\end{array}
\end{eqnarray}
where $p^{(\xi,\rho)}_{jk}$ are probabilities for the probe $\xi$ or signal $\rho$. Such measurements under a finite number of signal and probe copies result in frequency distributions $f^{(\rho)}_{jk}$ and $f^{(\xi)}_{jk}$ (see Fig.~\ref{fig:scheme}), which represent the data patterns for the signal state and the probe states, respectively.
So from Eqs.~\eqref{eq:expansion} and~\eqref{eq:freqprob} one has that the representation coefficients $x_{\xi}$ can be found by fitting the signal data pattern with probe data patterns:
\begin{equation}
\label{eq:freq}
f^{(\rho)}_{jk}\approx\sum_{\xi}x_{\xi}f^{(\xi)}_{jk},
\end{equation}
taking into account the physical constraints imposed on the density operator $\rho^{Appr}$ [$\rho^{Appr}=(\rho^{Appr})^{\dagger}$, $\Tr\rho^{Appr}=1$, $\rho^{Appr}\geq0$]. These constraints imply fulfillment of the following conditions for estimated coefficients:
\begin{equation}
\label{eq:constraints}
x_\xi=x_\xi^*,~\sum_{\xi}x_\xi=1,~\sum_{\xi}x_{\xi}\sigma_\xi\geq0.
\end{equation}

The possibility of accurate data-pattern reconstruction is closely related to the representation~\eqref{eq:expansion}. Having preliminary information or making a reasonable guess about the class of plausible signal states (such as the upper limit on the average photon number) allows one to make the reconstruction using an appropriately chosen set of probe states spanning required subspace. The problem of the optimal basis-state selection naturally arises. To solve this problem our research presented here follows certain selection criteria imposed on the basis states $\{\sigma_\xi\}$.

First of all, for practical purposes, it is advisable to use probe states that can be easily generated in the laboratory and provide an accurate representation of an unknown signal state. It was shown in previous works on the data-pattern scheme in the single-mode case that the usual coherent states satisfy these practical requirements rather well for a wide class of signal states~\cite{mogilevtsev-2010qt,mogilevtsev-2013}. Notice that some time ago considerable attention was paid to representing nonclassical quantum states in terms of the nonorthogonal basis of pure quantum states (here one can mention, for example, classical works \cite{jansky1,jansky2}). More recently, a number of works on representing entangled states using such coherent bases has appeared (for example, \cite{mikhalychev1,horoshko}). In the current work, we are implementing a quite different approach: a representation of the density matrix of the signal in terms of coherent-state projectors.

Second, it is desirable to use the smallest possible number of probe states in order to minimize computational resources for reconstruction of an unknown signal state. In subsequent research we use the fidelity $F(\rho,\rho^{Appr})=\sqrt{\rho^{1/2}\rho^{Appr}\rho^{1/2}}$ as the measure of quality of representation~\eqref{eq:expansion}~\cite{jozsa-1994}. Aiming for practical applications, we require the accuracy of representation~\eqref{eq:expansion} to  be greater than experimental measurement precision. It is not possible to evaluate the required accuracy precisely, but errors in recent works~\cite{cooper-2014,harder-2014} on the data-pattern scheme may be taken as the reference point for our estimations. Having relative experimental errors on the scale of several percent, we determine the required precision of the expansion~\eqref{eq:expansion} to be an order of magnitude greater. Thus, the criterion of the required accuracy of the representation can be expressed as $F(\rho,\rho^{Appr})\geq0.999$.
Notice that states that are close in terms of fidelity may have rather different physical properties, as has been shown theoretically and
experimentally, for example, in~\cite{paris1,paris2,paris3}. So, when judging the quality of the state representation, we estimate  also purity and the entanglement.

Third, the set of basis states must be suitable for representation of a wide range of quantum states with required precision. In this paper, we consider the broad class of states with small average number of photons which are widely used and applied in quantum cryptography and quantum computing~\cite{nielsen_chuang,gisin-2002}.

Next, we shall analyze the optimal basis sets of coherent projectors for the single-mode case as the initial simple problem. After that, based on the results obtained, we shall consider the case of entangled double-mode states.

\subsection{\label{sec:basis:singlemode}Single-mode case}

First, we consider the selection of the optimal basis sets for the single-mode case as the initial  problem. For this case we analyze the expansion for the single-photon state, the coherent state with amplitude $\alpha=0.5$, the even coherent state (the so-called ``Schr\"odinger's kitten'' state) ${\psi\propto{}|\alpha=0.5\rangle+|\alpha=-0.5\rangle}$, and the superposition of the vacuum and the single-photon states.
We assume the basis sets $\{\sigma_\xi=|\alpha_\xi\rangle\langle\alpha_\xi|\}$, where $\alpha_\xi$ are amplitudes of coherent projectors.
To solve the formulated problem we use \textsc{cvx} for \textsc{matlab}, a package for specifying and solving convex programs~\cite{cvx,boyd-2008}. The disciplined convex programming methodology is implemented in this system. It is assumed that one follows certain rules specifying a problem. If we take into account joint concavity of the fidelity $F$ in two arguments~\cite{nielsen_chuang}, it can be verified that the problem of expansion formulated above satisfies the whole set of rules. To improve convergence, we impose additional constraints on the absolute values of the coefficients $x_\xi$: $|x_\xi|\leq{}\Const$. It is most natural to select the discrete sets of coherent states $\{\sigma_\xi=|\alpha_\xi\rangle\langle\alpha_\xi|\}$ by constructing \textit{the square lattice} near the origin on the phase plane with the axes representing the values of the real and imaginary parts of complex amplitudes $\{\alpha_\xi\}$. Let us discuss now the possibility of representing single-mode quantum states based on these states and find the optimal sets, which meet the above-mentioned criteria. The optimization parameters of these basis sets are the number of nodes along each axis $N$ and grid pitch $d$.

\begin{figure}[!t]
\begin{minipage}[t]{8.6cm}
    \center{\includegraphics[width=7cm]{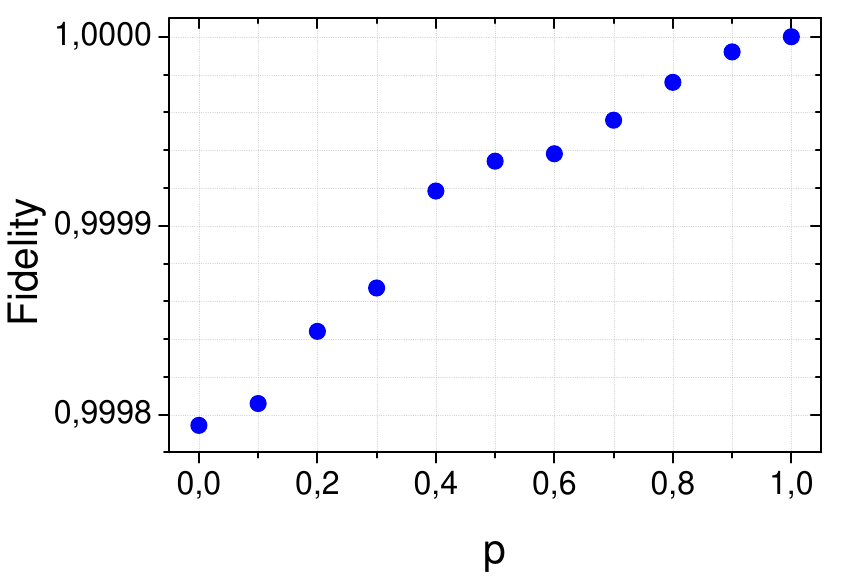} \\ \small{(a)}}
\end{minipage}
\hfill
\begin{minipage}[t]{8.6cm}
    \center{\includegraphics[width=7cm]{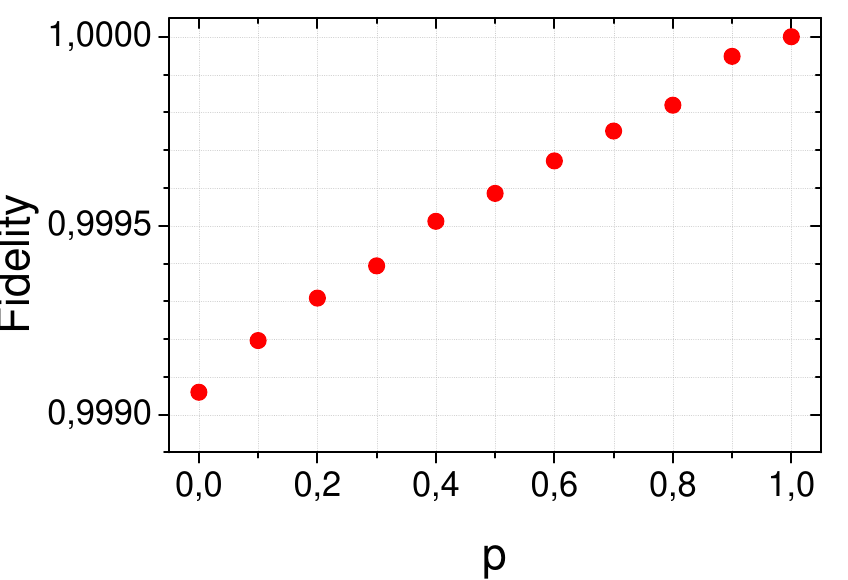} \\ \small{(b)}}
\end{minipage}
\caption{The fidelities of the representation of mixed states ${\rho=p|0\rangle\langle{}0|+(1-p)|1\rangle\langle{}1|}$ ($p\in[0,1]$) using coherent projectors with amplitudes selected in (a) the square lattice on the phase plain with $N\times{}N=6\times{}6$, $d=0.1$ and (b) the helical grid with $N=17$, $\Delta{}r=0.016$, $\Delta{}\varphi=\pi/4$ (b).}\label{fig:mixed-single}
\end{figure}

The increase in the accuracy of representation is provided by the increase in the number of nodes $N\times{N}$ (the number of coherent projectors) and the decrease in grid pitch $d$ since the recognition of small-scale details on the phase plane requires advanced resolution. On the other hand, filling a large area of the phase plain with the lattice with small $d$ requires an excessive number of basis states. This means that there are some optimal values of the number of nodes along each axis $N$ and grid pitch $d$. The dependencies of the fidelity on the parameters $N$ and $d$ enable us to find the optimal parameters of the square lattice taking into account the criterion of the required accuracy of the representation ($F\geq0.999$). The analysis of the representation~\eqref{eq:expansion} for the above-mentioned signal states enables us to determine optimal parameters in this case: $N\times{}N=6\times{}6$, $d=0.05-0.15$.

Notice that choosing probes not on the simple square grid but in a more sophisticated way may, in fact, lead to better accuracy with a smaller number of probes. For the signal with a small average number of photons we also consider the sets of coherent states chosen in the helical grid on the phase plane \cite{mogilevtsev-2014}. All probe states are equidistant in radius and angle; we optimize the number of nodes $N$, the step of the radial distance $\Delta{r}$, and the step of the angle $\Delta\varphi=2\pi(m/n)$, with $m,n\in{\mathds{N}}$.
The optimization procedure for this set of probes gives the following optimal parameters: $N=17$, $\Delta\varphi=\pi/4$, $\Delta{r}=0.009-0.016$. One can see that the use of the helical grid with optimal parameters requires indeed a smaller number of states in comparison with the square lattice.

The analysis of the representation~\eqref{eq:expansion} given above is for pure states. However, obviously,  the expansion~\eqref{eq:expansion} holds for mixed states as well. Let us demonstrate how the accuracy of the representation fares with the mixed states. For this purpose, we analyze the fidelity of the representation for the mixed states ${\rho=p|0\rangle\langle{}0|+(1-p)|1\rangle\langle{}1|}$ ($p\in[0,1]$). Figure~\ref{fig:mixed-single} shows that the fidelities of the representation of mixed states exceed the minimum required accuracy for optimal basis sets of coherent projectors in the cases of the square lattice and the helical grid. This analysis confirms that the method considered works appropriately for mixed states as well.

\subsection{\label{sec:basis:doublemode}Double-mode case}
\begin{figure}[t]
\includegraphics[width=8.6cm]{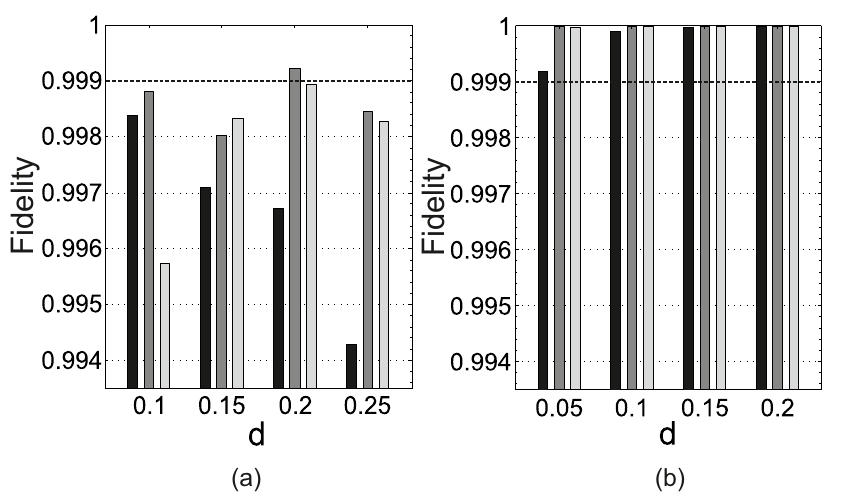}
\caption{The fidelity of the representation of entangled double-mode states against the grid pitch $d$ of the square lattice of coherent projectors: (a) the number of nodes along each axis $N=6$; (b) the number of nodes along each axis $N=7$. The predefined double-mode states are ${\psi=\Const\times\big(|\alpha\rangle_1|-\alpha\rangle_2+|-\alpha\rangle_1|\alpha\rangle_2\big)}$ with $\alpha=0.5$ (black bars), ${\psi=\big(|0\rangle_1|1\rangle_2+|1\rangle_1|0\rangle_2\big)/\sqrt{2}}$ (dark gray bars), ${\psi=\big(|0\rangle_1|0\rangle_2+|1\rangle_1|1\rangle_2\big)/\sqrt{2}}$ (light gray bars).}\label{fig:fidel-double}
\end{figure}

Now let us move to the consideration of the multimode case. Since we are aiming, primarily, to show how the entanglement survives the data-pattern inference, we restrict ourselves to the double-mode case. The representations are analyzed for the following states: ${\psi=\big(|0\rangle_1|1\rangle_2+|1\rangle_1|0\rangle_2\big)/\sqrt{2}}$, ${\psi=\big(|0\rangle_1|0\rangle_2+|1\rangle_1|1\rangle_2\big)/\sqrt{2}}$, $\psi=\Const\times\big(|\alpha\rangle_1|-\alpha\rangle_{2}+ |-\alpha\rangle_1|\alpha\rangle_2\big)$ (${\alpha=0.5}$).
We choose the basis sets ${\{\sigma_\xi=|\alpha_{\xi_1}\rangle_1\langle\alpha_{\xi_1}|\otimes|\alpha_{\xi_2}\rangle_2\langle\alpha_{\xi_2}|\}}$ as the tensor product of the coherent states selected in the nodes of the square lattice on the phase plane. The parameters to optimize are the number of nodes along each axis $N$ and grid pitch $d$ for the two modes.

Accurate representation~\eqref{eq:expansion} of states that are the tensor product of pure single-mode states is possible for the parameters found in the previous section ($N=6$, $d=0.05-0.15$) since ${F(\rho_1\otimes\rho_2,\rho_1^{Appr}\otimes\rho_2^{Appr})}={F(\rho_1,\rho_1^{Appr})F(\rho_2,\rho_2^{Appr})}$~\cite{jozsa-1994}. Intuitively, in the two-mode case it is natural to expect some nontrivial complication due to the presence of entanglement. We analyze the expansions~\eqref{eq:expansion} for the pure entangled states mentioned above in this section. The results presented in Fig.~\ref{fig:fidel-double} indicate that, indeed, the representation of entangled double-mode states requires a greater number of nodes along each axis $N$ than in the single-mode case. This figure demonstrates that a representation with fidelity $F\geq0.999$ is possible for the sets with the following optimal parameters: $N=7$, $d=0.05-0.20$.

\begin{figure}[t]
\includegraphics[width=7.6cm]{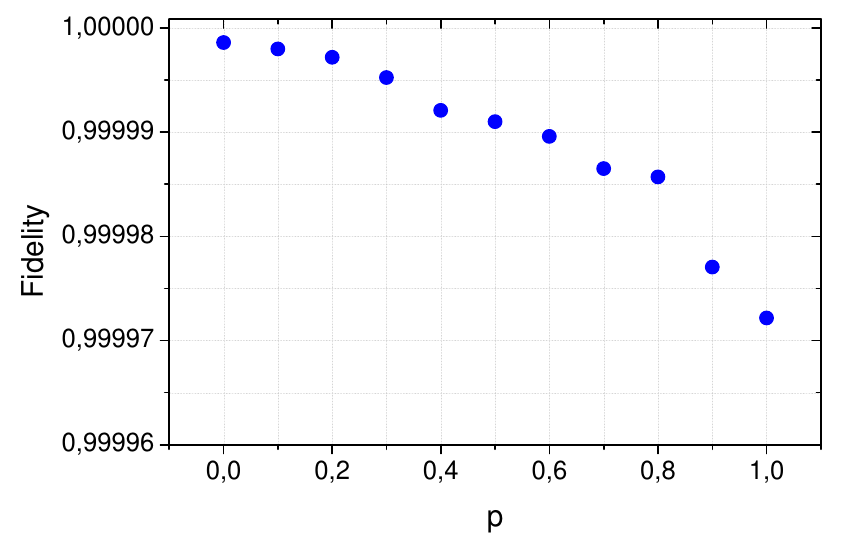}
\caption{The fidelities of the representation of entangled mixed states ${\rho=\frac{1-p}{2}\big(|0\rangle_1|1\rangle_2+|1\rangle_1|0\rangle_2\big)\big(\langle0|_1\langle1|_2+\langle1|_1\langle0|_2\big)}+{\frac{p}{2}\big(|0\rangle_1|0\rangle_2+|1\rangle_1|1\rangle_2\big)\big(\langle0|_1\langle0|_2+\langle1|_1\langle1|_2\big)}$ ($p\in[0,1]$) using coherent projectors with amplitudes selected in the square lattice on the phase plain with $N=7$, $d=0.05$.}\label{fig:mixed-double}
\end{figure}

It is also useful to demonstrate that the method considered works properly for mixed states in this case. For this reason, we consider the expansion of the state with the density operator ${\rho=\frac{1-p}{2}\big(|0\rangle_1|1\rangle_2+|1\rangle_1|0\rangle_2\big)\big(\langle0|_1\langle1|_2+\langle1|_1\langle0|_2\big)}\hmm+{\frac{p}{2}\big(|0\rangle_1|0\rangle_2+|1\rangle_1|1\rangle_2\big)\big(\langle0|_1\langle0|_2+\langle1|_1\langle1|_2\big)}$ ($p\in[0,1]$). An analysis of the fidelities of the representation of these states confirms that optimal basis sets of coherent projectors are appropriate for the accurate expansion of mixed states as well (see Fig.~\ref{fig:mixed-double}).

We close this section by noting that the basis sets of coherent projectors found are applicable for the essentially accurate representation of single-mode and entangled double-mode states with a small average number of photons. We note that the achieved precision of the representation~\eqref{eq:expansion}, $F(\rho,\rho^{Appr})\geq0.999$, indicates that these optimal sets of probe states can reliably represent a wide class of signal states. This circumstance allows us to expect that these sets of coherent projectors may be successfully implemented in data-pattern tomography of single-mode and entangled double-mode states effectively.

\section{\label{sec:analysis}Expansion analysis}
In this section we consider the quality of the expansion~\eqref{eq:expansion} using discrete sets of coherent projectors. Since any real measurements are connected with the noise of different sources, it is essential to analyze the stability of the representation using the basis sets considered. After that, we shall analyze the conservation of entanglement for the expansion based on the optimal discrete sets. Last, the purity of the expansions is investigated based on the analysis of their eigenvalues.

\subsection{\label{sec:analysis:stability}Stability}

\begin{figure}[t]
\begin{minipage}[t]{4.25cm}
    \center{\includegraphics[width=4.25cm]{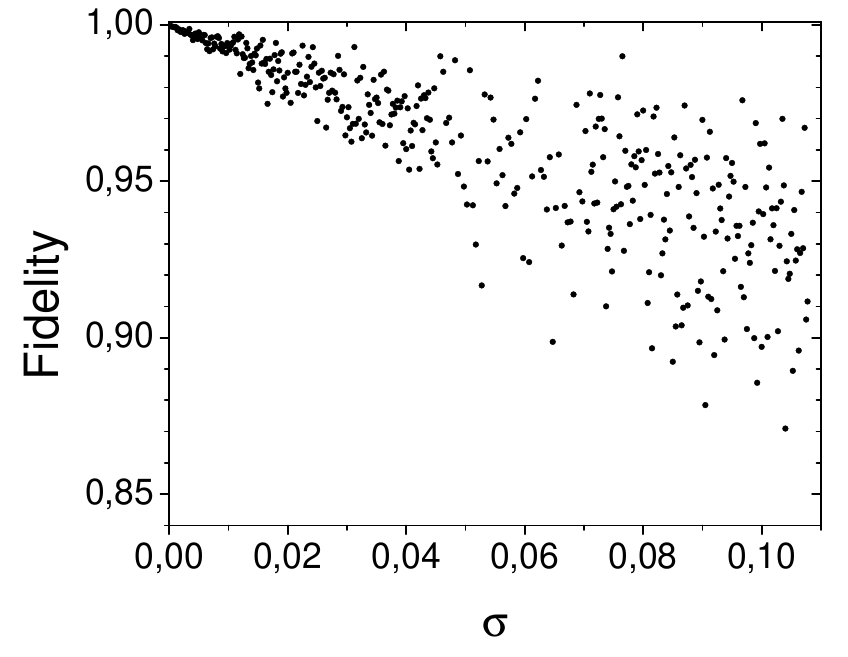} \small{(a)}}
\end{minipage}
\hfill
\begin{minipage}[t]{4.25cm}
    \center{\includegraphics[width=4.25cm]{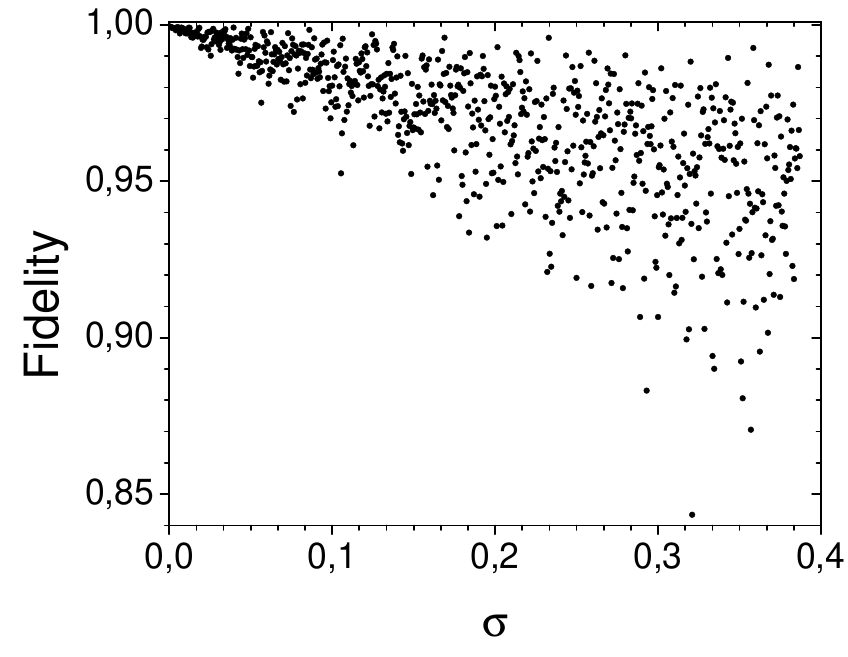} \small{(b)}}
\end{minipage}
\begin{minipage}[t]{4.25cm}
    \center{\includegraphics[width=4.25cm]{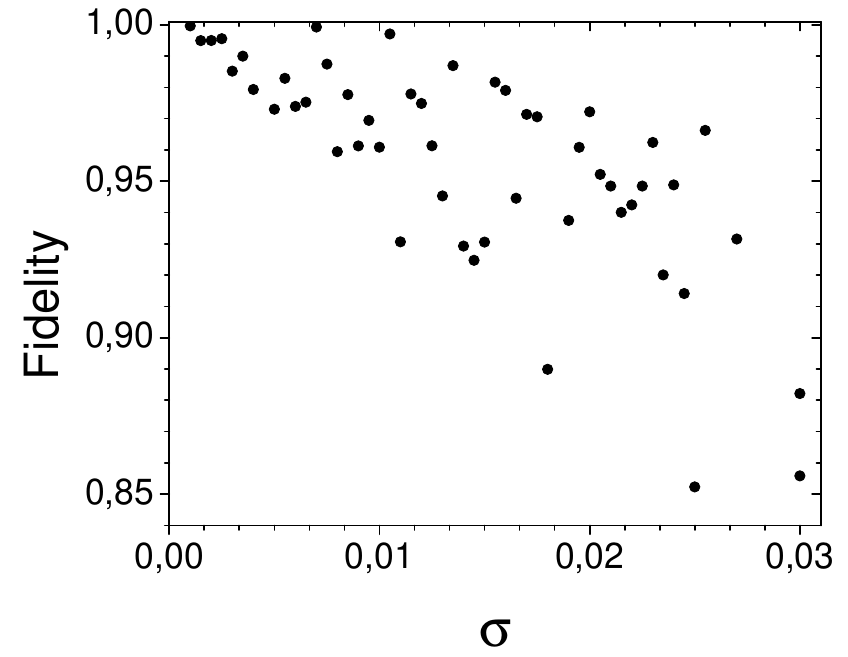} \small{(c)}}
\end{minipage}
\caption{The fidelities of the representation~\eqref{eq:expansionfluct} against the rms amplitude error. (a) The fidelities of the representation of the coherent state with amplitude $\alpha=0.5$ for the square lattice of coherent projectors with $N=6$, $d=0.15$. (b) The fidelities of the representation of coherent state with amplitude $\alpha=0.5$ for the helical grid with $N=17$, $\Delta{}r=0.016$, $\Delta{}\varphi=\pi/4$. (c) The fidelities of the representation of the state ${\psi=\big(|0\rangle_1|1\rangle_2+|1\rangle_1|0\rangle_2\big)/\sqrt{2}}$ using coherent projectors in the pitches of the square lattice with $N=7$, $d=0.15$.}\label{fig:noise}
\end{figure}
Analyzing the stability of the reconstructions using the optimal basis sets, we compare quantum state $\rho$ with the result of the expansion containing fluctuations with normal distribution $N(0,\sigma^2)$ in the coefficients:
\begin{equation}
\label{eq:expansionfluct}
\rho^{Appr\prime}=\sum_{\xi=1}^{N}(x_\xi+\{\text{noise}\})\sigma_\xi.
\end{equation}
Notice that we enforce semipositivity and unit trace of $\rho^{Appr\prime}$. Plots of the fidelity $F\big(\rho,\rho^{Appr\prime}\big)$ against the rms amplitude error indicate that the representations using the optimal sets of coherent states are quite stable for both the single- and double-mode states. The plots for the coherent state with amplitude $\alpha=0.5$ in the single-mode case and for state ${\psi=\big(|0\rangle_1|1\rangle_2+|1\rangle_1|0\rangle_2\big)/\sqrt{2}}$ in the double-mode case are presented in Fig.~\ref{fig:noise}. One can see that the  fidelity remains rather high over a quite large range of the rms amplitude error.

\subsection{\label{sec:analysis:entanglement}Entanglement estimation}
It is not obvious that the expansion~\eqref{eq:expansion} based on the set of coherent projectors conserves the entanglement. In order to demonstrate this, we estimate entanglement for the same states represented using the square lattice of basis states.

\begin{figure}[!t]
\begin{minipage}[t]{8.6cm}
    \center{\includegraphics[width=6cm]{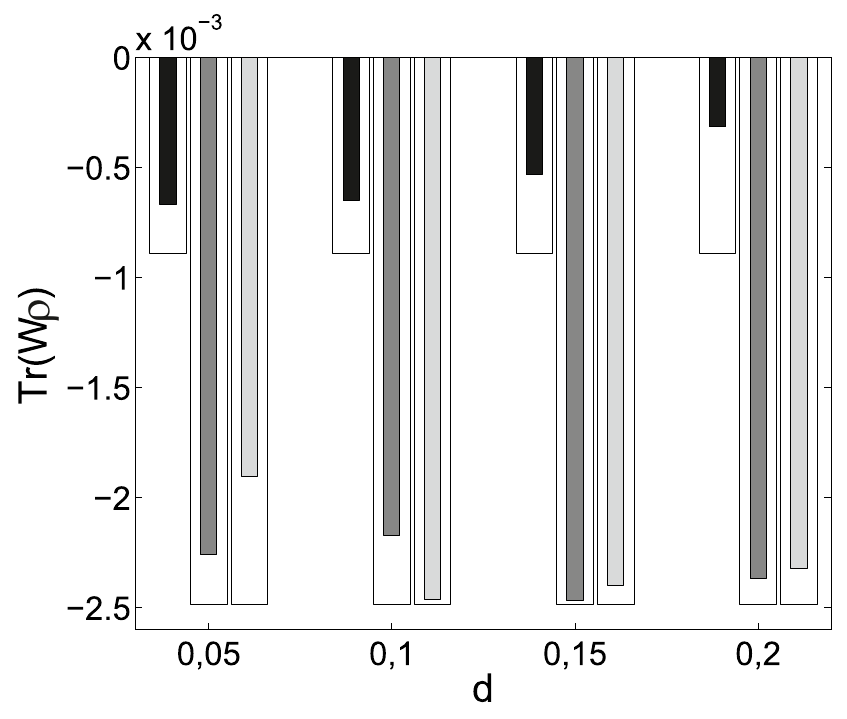} \\ \small{(a)}}
\end{minipage}
\hfill
\begin{minipage}[t]{8.6cm}
    \center{\includegraphics[width=6cm]{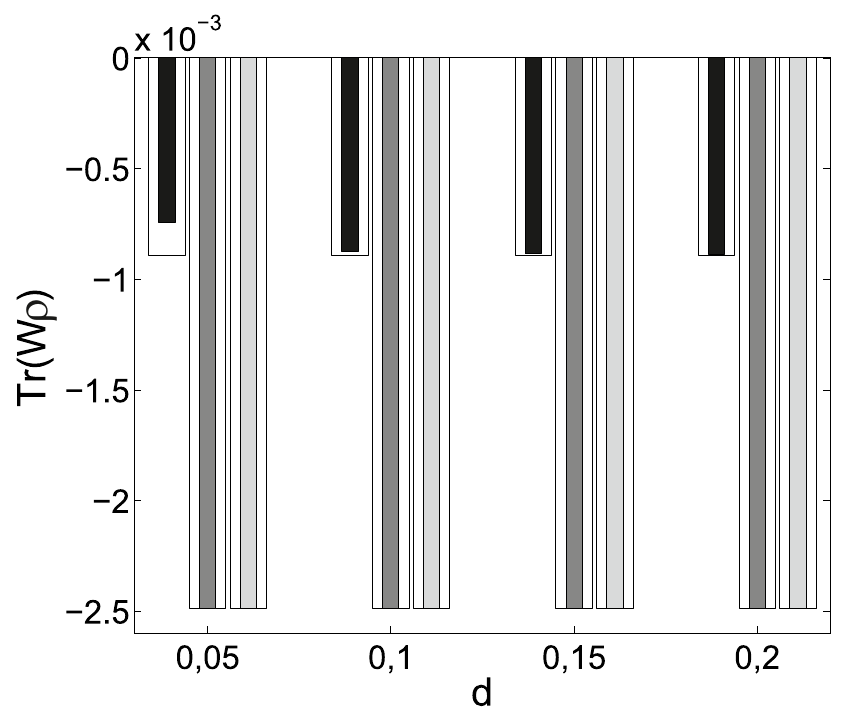} \\ \small{(b)}}
\end{minipage}
\caption{Plots of the trace $\Tr{(W\rho)}$ against the grid pitch $d$ of the square lattice on the phase plane: (a) the number of nodes along each axis $N=6$ (fidelity level of representation $\rho^{Appr}$ is $0.994-0.999$); (b) the number of nodes along each axis $N=7$ (fidelity level of representation $\rho^{Appr}$ is $0.999-0.99999$). The entanglement witnesses $W$ are obtained solving the optimization problem for the precise density matrices. Filled bars represent the results for the density operators $\rho^{Appr}$ being the representation of the following states: ${\psi=\Const\times\big(|\alpha\rangle_1|-\alpha\rangle_2+|-\alpha\rangle_1|\alpha\rangle_2\big)}$ with $\alpha=0.5$ (black bars), ${\psi=\big(|0\rangle_1|1\rangle_2+|1\rangle_1|0\rangle_2\big)/\sqrt{2}}$ (dark gray bars) and ${\psi=\big(|0\rangle_1|0\rangle_2+|1\rangle_1|1\rangle_2\big)/\sqrt{2}}$ (light gray bars). White bars represent the trace $\Tr{(W\rho)}$ for the precise density matrices of the corresponding states.}\label{fig:EW}
\end{figure}

We use an entanglement witness (EW) to determine whether a state is separable or not. A density operator $\rho$ describes an entangled state iff there exists a Hermitian operator $W$ (called EW) which detects its entanglement, i.e., $\Tr(W\rho)<0$ and $\Tr(W\sigma_{sep})>0$ for all $\sigma_{sep}$ separable~\cite{Horodecki1996,brandao-2006}. We calculate the entanglement witness using a method proposed in Ref.~\cite{brandao-2004}. For the calculation of the operator $W$ we use \textsc{cvx}~\cite{cvx,boyd-2008}. Figure~\ref{fig:EW} shows the results of these calculations for the square lattice of basis states with $N=6$ and $N=7$. One can see that all values of the trace $\Tr(W\rho)$ estimated are negative. Thus, the representation~\eqref{eq:expansion} with these discrete sets of coherent projectors conserves entanglement. Figure~\ref{fig:EW}(b) demonstrates that the values of the trace $\Tr{(W\rho^{Appr})}$ for the expansions using optimal basis sets are very close to the values found for the precise density matrices, which indicates the closeness of their entanglements.

\subsection{\label{sec:analysis:purity}Purity}

\begin{figure}[t]
\includegraphics[width=8.6cm]{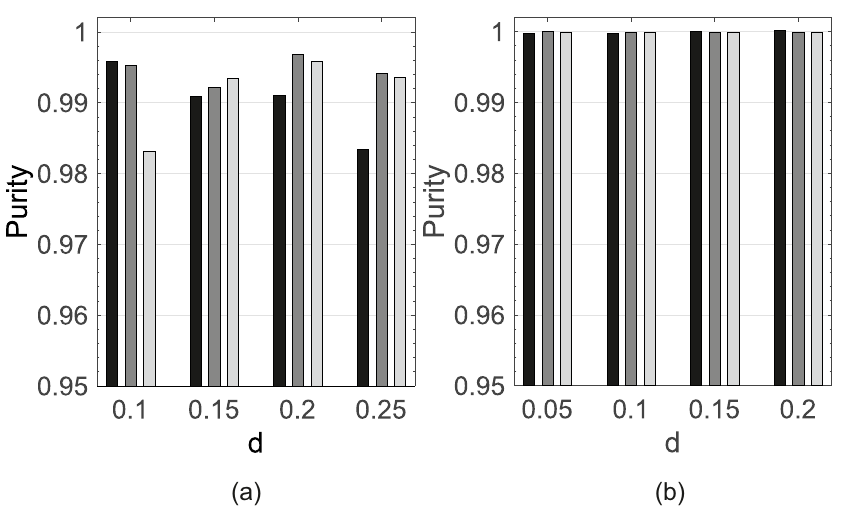}
\caption{The purity of the representation of entangled double-mode states against the grid pitch $d$ of the square lattice of coherent projectors: (a) the number of nodes along each axis $N=6$ (fidelity level of representation $\rho^{Appr}$ is $0.994-0.999$); (b) the number of nodes along each axis $N=7$ (fidelity level of representation $\rho^{Appr}$ is $0.999-0.99999$). The states considered are ${\psi=\Const\times\big(|\alpha\rangle_1|-\alpha\rangle_2+|-\alpha\rangle_1|\alpha\rangle_2\big)}$ with $\alpha=0.5$ (black bars), ${\psi=\big(|0\rangle_1|1\rangle_2+|1\rangle_1|0\rangle_2\big)/\sqrt{2}}$ (dark gray bars), ${\psi=\big(|0\rangle_1|0\rangle_2+|1\rangle_1|1\rangle_2\big)/\sqrt{2}}$ (light gray bars).}\label{fig:purity_entangled}
\end{figure}

At the end of this section we analyze the purity of the density matrices $\rho^{Appr}$, which is defined as ${\mu[\rho^{Appr}]=\Tr{(\rho^{Appr})^2}=\sum_{k}\lambda_{k}^2}$ ($\lambda_{k}$ are the eigenvalues of the density matrix $\rho^{Appr}$). We analyze the purity of the density matrices $\rho^{Appr}$ of the states considered earlier. The calculations for the expansions using the optimal basis sets show that the purity is well conserved in the single-mode case. Figure~\ref{fig:purity_entangled} shows the dependence of the purity of the representation of entangled double-mode states against the grid pitch $d$ of the square lattice of coherent projectors with $N=6$ and $N=7$. According to Fig.~\ref{fig:purity_entangled}(b) the expansions using the optimal basis sets are found to conserve the purity with great precision.

In conclusion, we can assert that the optimal basis sets considered in this section are applicable for a highly accurate representation of entangled states with a small average number of photons.

\section{\label{sec:reconstruction}Reconstruction using the optimal basis sets}
In this section, we demonstrate the possibility of accurate data-pattern reconstruction of single-mode and entangled double-mode states using the optimal basis sets of the coherent projectors found. To this end, we shall specify and simulate the set of measurements for this scheme.

\subsection{\label{sec:reconstruction:measurments}Set of measurements}
As a means of demonstrating the applicability of the discrete basis sets of coherent projectors in data-pattern tomography in the simplest and most straightforward way, let us take the intended measurements to be projections onto coherent states for single- and double-mode cases.  These measurements are described by the POVM elements $\Pi_{j}=|\beta_j\rangle\langle\beta_j|$ and $\Pi_j=|\beta_{j_1}\rangle_j\langle\beta_{j_1}|\otimes|\beta_{j_2}\rangle_j\langle\beta_{j_2}|$ ($j=1,\dots{},K$)  with amplitudes $\beta_j$, $\beta_{j_1}$, $\beta_{j_2}$. For the assumed ideal lossless detection, the probabilities $p^{(\xi)}_{j}$ and $p^{(\rho)}_{j}$ of observing the positive outcome in the $j$th measurement for a probe state $\sigma_{\xi}$ and a signal state $\rho$ are given by Eqs. (\ref{eq:freqprob}).
Amplitudes of the coherent projectors forming the POVM elements $\{\Pi_j\}$ are selected as equidistant phase-space points that form a square lattice in our simulations. We assume that the probabilities of observing every coherent-state setting $j$ for the probe state $\sigma_{\xi}$ and the signal state $\rho$ are measured with a finite number of state copies $N_{rep}$. The experimental frequencies $f^{(\xi)}_{f}$ and $f^{(\rho)}_{f}$ are simulated using a binomial distribution with parameters $N_{rep}$, $p^{(\xi)}_{j}$ and $N_{rep}$, $p^{(\rho)}_{j}$ for the sets of probe states and the signal state, respectively.

One is able to carry out the reconstruction minimizing the distance
\begin{equation}
\label{eq:functional}
E[x_\xi]=\sum_{j}\Big(f^{(\rho)}_{j}-\sum_{\xi}x_{\xi}f^{(\xi)}_{j}\Big)^2.
\end{equation}
Minimization of the functional~\eqref{eq:functional} subject to the constraints imposed on the coefficients $\{x_{\xi}\}$ represents the semidefinite convex problem~\cite{boyd-2004} and can be solved using the package \textsc{cvx}~\cite{cvx,boyd-2008}.
\begin{figure}[!t]
\includegraphics[width=8.6cm]{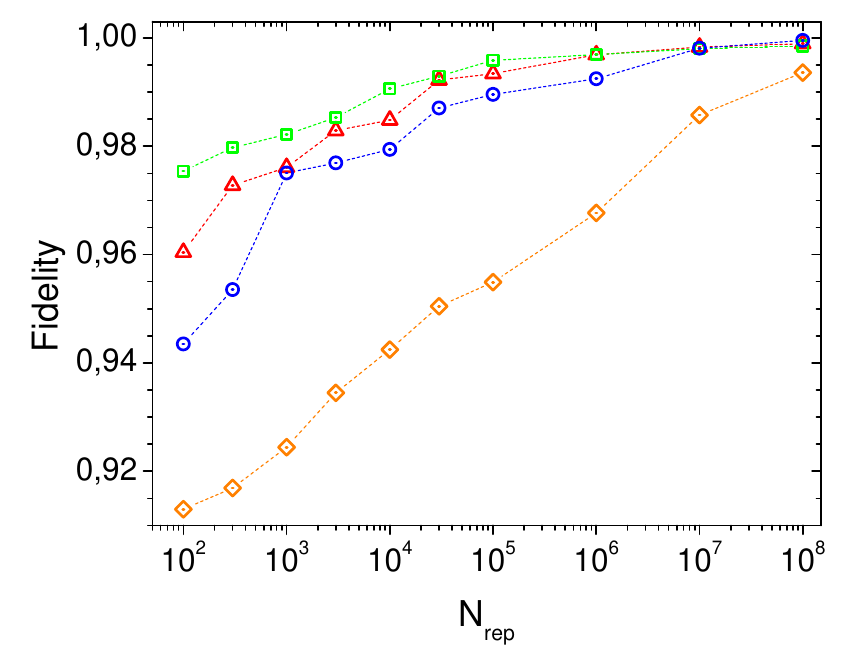}
\caption{The fidelities of the reconstruction of single-mode states against the number of state copies $N_{rep}$. The amplitudes of the probe states and projectors $\{\Pi_j\}$ are chosen in the nodes of the square lattice with the optimal parameters $N=6$, $d=0.15$. The reconstruction process is simulated for the single-photon state (rhombuses), the coherent state with amplitude $\alpha=0.5$ (squares), the even coherent state with amplitude ${\alpha=0.5}$ (triangles), and the superposition of the vacuum and the single-photon states (circles).}\label{fig:reconstruction-single}
\end{figure}

\subsection{\label{sec:reconstruction:results}Results}
We are now ready to demonstrate the possibility of accurate data-pattern reconstruction using the discussed optimal sets for the reconstruction of single-mode and entangled double-mode states. Typically, the number of employed measurement settings $K$ is not equal to the number of probe states $M$. However, for the sake of simplicity we let $K=M$ and take the square lattice for the POVM elements $\{\Pi_j\}$ to be the same as that for the set of basis states $\{\sigma_{\xi}\}$.

\begin{figure}[!t]
\includegraphics[width=8.6cm]{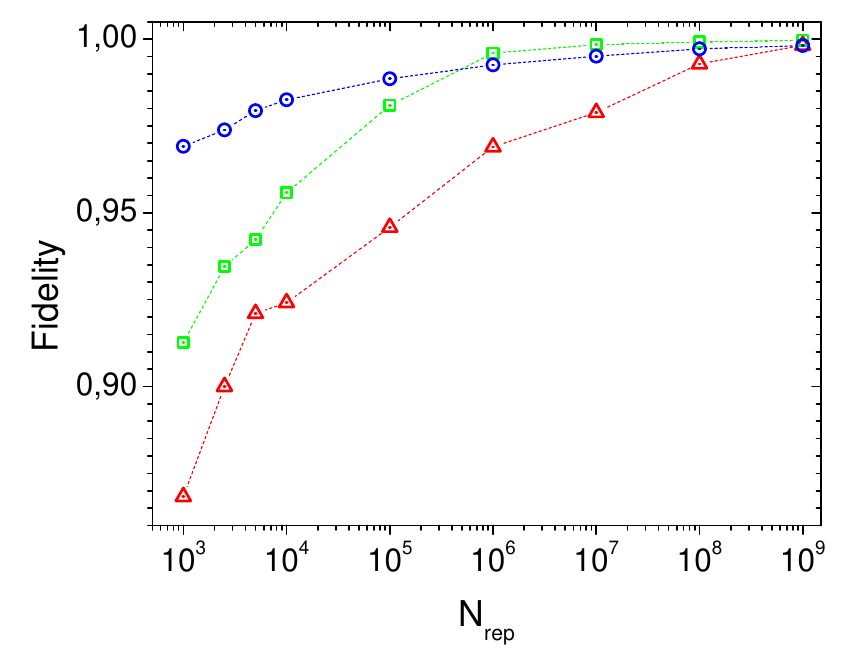}
\caption{The fidelities of the reconstruction of entangled double-mode states against the number of state copies $N_{rep}$. The amplitudes of the probe states and projectors $\{\Pi_j\}$ are chosen in the nodes of the square lattice with the optimal parameters $N=7$, $d=0.15$. The reconstruction is done for ${\psi=\Const\times\big(|\alpha\rangle_1|-\alpha\rangle_2+|-\alpha\rangle_1|\alpha\rangle_2\big)}$ with $\alpha=0.5$ (circles), ${\psi=\big(|0\rangle_1|1\rangle_2+|1\rangle_1|0\rangle_2\big)/\sqrt{2}}$ (squares), ${\psi=\big(|0\rangle_1|0\rangle_2+|1\rangle_1|1\rangle_2\big)/\sqrt{2}}$ (triangles).}\label{fig:reconstruction-double}
\end{figure}

In the single-mode case we select the amplitudes of probe states and projectors $\{\Pi_j=|\beta_j\rangle\langle\beta_j|\}$ to be in the nodes of the square lattice with the optimal parameters $N=6$, $d=0.15$.  Figure~\ref{fig:reconstruction-single} shows examples of fidelities of the single-mode states reconstructed against the number of state copies $N_{rep}$. In the same way we carry out the reconstruction of entangled double-mode states by selecting the amplitudes of the discrete sets of probe states and projectors $\{\Pi_j=|\beta_{j_1}\rangle_j\langle\beta_{j_1}|\otimes|\beta_{j_2}\rangle_j\langle\beta_{j_2}|\}$ as phase-space points that form the square lattice with optimal parameters $N=7$, $d=0.15$. Figure~\ref{fig:reconstruction-double} demonstrates the fidelities of this reconstruction process for entangled double-mode states. Figures~\ref{fig:reconstruction-single} and~\ref{fig:reconstruction-double} show that the fidelities $F(\rho,\rho^{Appr})$ for single-mode and entangled double-mode states are comparably close for reasonably large numbers of state copies ($N_{rep}>10^4$). One can see that an observer is able to reconstruct single-mode or entangled double-mode states with a fidelity that is arbitrarily close to the target fidelity $F(\rho,\rho^{Appr})=0.999$ by increasing the number of state copies $N_{rep}$. So for sufficiently high accuracy of the representation the fidelity of the signal reconstructions for a large number of signal state copies indeed tends to the values close to the fidelity given by the representation. Furthermore, this number of copies stands well within the region of experimental feasibility.

\begin{figure}[!t]
\includegraphics[width=8.6cm]{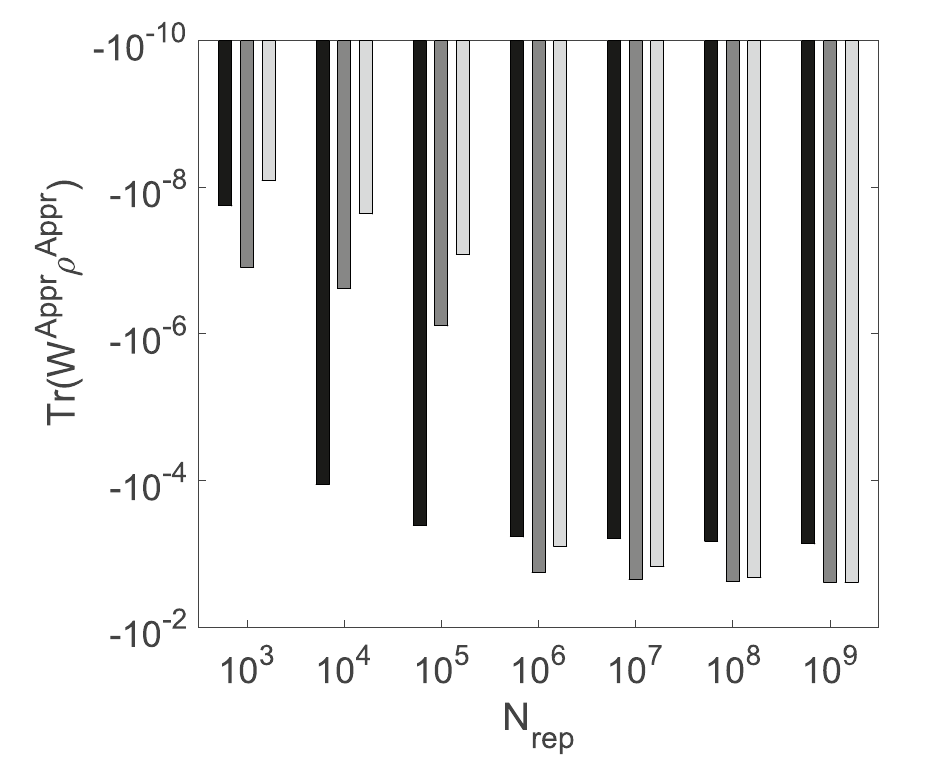}
\caption{Plot of the trace $\Tr{(W^{Appr}\rho^{Appr})}$ against the number of state copies $N_{rep}$ for the following signal states: ${\psi=\Const\times\big(|\alpha\rangle_1|-\alpha\rangle_2+|-\alpha\rangle_1|\alpha\rangle_2\big)}$ with $\alpha=0.5$ (black bars), ${\psi=\big(|0\rangle_1|1\rangle_2+|1\rangle_1|0\rangle_2\big)/\sqrt{2}}$ (dark gray bars), and ${\psi=\big(|0\rangle_1|0\rangle_2+|1\rangle_1|1\rangle_2\big)/\sqrt{2}}$ (light gray bars). The entanglement witnesses $W^{Appr}$ are obtained by solving the optimization problem for the reconstructed density matrices. The values of $\Tr{(W\rho)}$ by obtained solving the optimization problem for the precise density matrices are approximately equal to $-0.00089$ for the signal state  ${\psi=\Const\times\big(|\alpha\rangle_1|-\alpha\rangle_2+|-\alpha\rangle_1|\alpha\rangle_2\big)}$ and $-0.00249$ for the signal states ${\psi=\big(|0\rangle_1|1\rangle_2+|1\rangle_1|0\rangle_2\big)/\sqrt{2}}$ and ${\psi=\big(|0\rangle_1|0\rangle_2+|1\rangle_1|1\rangle_2\big)/\sqrt{2}}$. The fidelity level for the reconstructed states is $0.86-0.999$.}\label{fig:EW-reconstruction}
\end{figure}

Finally, we analyze the conservation of entanglement for the signal states reconstructed with different numbers of state copies $N_{rep}$. For this purpose, we calculate the entanglement witness $W$ using a method proposed in Ref.~\cite{brandao-2004}. An entanglement witness operator calculated for the precise density matrix $\rho$ does not have to be EW for the states $\rho^{Appr}$ reconstructed with a relatively small number of state copies. Therefore, for the calculation of the entanglement witness $W$ we solve the problem of convex optimization for the density operators $\rho^{Appr}$ reconstructed. Figure~\ref{fig:EW-reconstruction} shows the results of these calculations for the same sets of probe states and projectors as in Fig.~\ref{fig:reconstruction-double} (we select them in the square lattice on the phase plane with parameters $N=7$, $d=0.15$). Figure~\ref{fig:EW-reconstruction} demonstrates that the entanglement survives the inference procedure even for a reasonably small number of state copies $N_{rep}$.

We conclude that using a discrete probe set of appropriately chosen coherent projectors in data-pattern tomography enables us to accurately reconstruct entangled optical quantum states.

\section{\label{sec:conclusions}Conclusions}
We have discussed the data-pattern approach to quantum tomography of entangled states. The efficiency of this procedure depends essentially on the choice of the basis set of the probe states. We substantiated the choice of the discrete set of coherent states by constructing a regular grid of basis states and finding the optimal expansion for given quantum states based on this basis set using the method considered in the paper. We have demonstrated the possibility of accurate representation of entangled states based on these discrete sets of coherent projectors, found the optimal ones for entangled double-mode states as well as single-mode states with a small average number of photons, and demonstrated the robustness of the representation with respect to added noise. The simulations of the reconstruction process demonstrated the feasibility and the effectiveness of data-pattern quantum tomography of entangled states using the discrete basis set of coherent states. The results presented show that the data-pattern approach may become an efficient tool in experimental quantum-state reconstruction of entangled states.\\[5pt]

\section*{ACKNOWLEDGMENTS}
This work was supported by
the National Academy of Sciences of Belarus through the
program “Convergence” and the European Commission through
the SUPERTWIN project (Contract No. 686731).

\end{document}